# Design of Partially Etched GaP-OI Microresonators for Two-Color Kerr Soliton Generation at NIR and MIR

Houling Ji[1], Zhaoting Geng[1], Weiren Cheng[1], Zhuoyu Yu[1], Pengzhuo Wu[1], Yi Li[1†] and Qiancheng Zhao[1*]

[1]*School of Microelectronics, Southern University of Science and Technology, Shenzhen, Guangdong, China, 518000*

e-mail: †liy37@sustech.edu.cn, *zhaoqc@sustech.edu.cn

*Abstract*—We present and theoretically investigate a dispersion engineered GaP-OI microresonator containing a partially-etched gap of 250 nm × 410 nm in a 600 nm × 2990 nm waveguide. This gap enables a 3.25 μm wide anomalous dispersion spectral span covering both the near-infrared and the mid-infrared spectra. This anomalous dispersion is manifested by two mechanisms, being the hybridization of the fundamental TE modes around 1550 nm and the geometric dispersion of the higher order TE mode around the 3100 nm wavelengths, respectively. Two Kerr soliton combs can be numerically generated with 101 GHz and 97 GHz teeth spacings at these spectral windows. The proposed structure demonstrates the design flexibility thanks to the partially etched gap and paves the way towards potential coherent multicolor frequency comb generation in the emerging GaP-OI platform.

## I. INTRODUCTION

Microresonator-based optical frequency combs (OFC) are widely used in microwave generation [1], optical atomic clocks [2], coherent optical telecommunications [3], chip-scale distance ranging [4], exoplanet exploration [5], and spectroscopy [6]. While numerous applications were demonstrated at near-infrared (NIR) regime ($\lambda$~0.7-2.5 μm), molecular spectroscopy particularly favors the mid-infrared (MIR) spectral range ($\lambda$~2.5-20 μm) since a large number of molecules exhibit strong vibrational fingerprints in this region [7]. Bridging multiple spectral windows together on one single comb resonator turns to be of utmost importance, extending the usable spectrum range and promoting interdisciplinary innovations. To this end, conventional octave-spanning [8] can blur the boundary between the NIR and MIR spectral regions, manifested by soliton states that possess broad spectral range with phase coherence. However, the soliton power drops as the comb line frequency moves away from the pump frequency, following the characteristic sech$^2$ shape, which limits the available power in either the NIR or the MIR spectral region. Dispersive waves [9] can mitigate the power inefficiency problem at spectrum tails, but their narrow bandwidth and sensitivity to fabrication variations pose other challenges. Instead of self-limiting with a single comb window, multi-color frequency comb generation with coherent pump sources provides an alternative solution to the dilemma between the bandwidth and the comb line power, without losing coherence [10]–[12].

The interest in achieving multiple OFCs on a single microresonator lies in but is not limited to the following reasons: (i) the comb line power in each spectral region is controlled by its individual pump source and can be adjusted independently; (ii) a broad frequency comb spectrum can be realized by combining several smaller frequency comb windows, which relaxes the waveguide geometric constraints to implement an extraordinarily wide anomalous dispersion profile; (iii) each sub frequency comb window can be optimized individually with unmatched design flexibility; (iv) it opens the possibility for Kerr comb interaction with other nonlinear effects such as second-harmonic generation, Raman scattering, or optical parametric oscillation to exhibit two or more frequency combs at a time with merely a single pump source [13].

Multicolor frequency combs have been investigated on SiC [10], Si$_3$N$_4$ [11], AlN [12], and LiNbO$_3$ [14] resonators. Recent studies turn to GaP integrated photonic platform [15]–[17]. GaP has a wide transparent window (0.55 μm – 11 μm) covering visible, NIR and MIR spectral ranges. Above 1.1 μm, two-photon absorption can be neglected which renders GaP low nonlinear loss at telecommunication bands. GaP has a conspicuous 3$^{rd}$ order nonlinearity with $n_2 = 11 \times 10^{-18}$ m$^2$W$^{-1}$, at 1.55 μm wavelength, favoring four-wave-mixing (FWM) process for Kerr frequency comb generation. In tandem with FWM, its large $\chi^{(2)}$ nonlinearity (82 pmV$^{-1}$ at $\lambda$~1.55 μm) enables frequency comb translation into other spectral windows via second harmonic generation or parametric down conversion process. The advantage of large refractive index ($n > 3$ at C-band) is further enhanced by the GaP on Insulator (GaP-OI) architecture [18]–[20], which enables tight mode confinement and improved fabrication compatibility. These merits make the GaP-OI integrated photonic platform an ideal candidate for frequency comb generation in multiple spectral windows.

Here, we present the design of a dispersion-engineered GaP-OI microresonator with a partially etched gap for two-color frequency comb generation at NIR and MIR spectra. The resonator has a radius of 140 μm and a cross-section of 600 nm × 2990 nm. By inserting a partially etched gap in the waveguide, anomalous dispersion can be individually realized around the 1.55 μm and the 3.1 μm pump wavelengths. The partially etched gap provides three more tuning knobs in dispersion engineering, being the gap width, gap depth, and the location of the gap. Through synergistic optimization of the gap as well as the waveguide dimension, we achieve an anomalous dispersion spectral span of 3.25 µm with a gap size of 250 nm ×410 nm, as compared to 3.05 µm without the gap. Kerr soliton combs are numerically investigated, covering from 1.12 µm to 2.38 µm over 0.72 dBm power level at 1.55 µm pump wavelength and from 2.35 µm to 4.73 µm over 12.3 dBm power level at 3.1 µm pump wavelength. This design paves the way towards the two-color coherent frequency comb generation at the NIR and MIR spectra using a single pump source, in combination with GaP's quadratic nonlinearity.

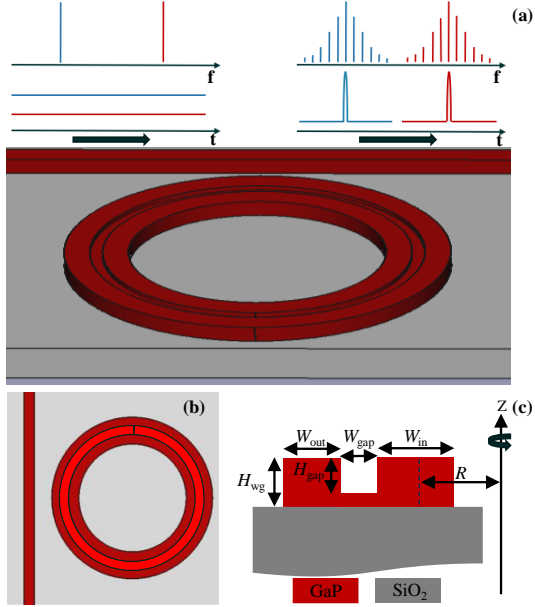

Figure 1 GaP trench ring design parameters on $SiO_2$ substrate with $SiO_2$ top cladding ($SiO_2$ upper cladding not shown for clarity). (a) Schematic of the GaP trench-ring geometry. (b) Top view, and (c) cross-section view. $H_{wg}$, $W_{in}$ and $W_{out}$ refer to the dimension of the GaP ring. $H_{gap}$ and $W_{gap}$ refer to the dimension of the trench.

## II. WAVEGUIDE DISPERSION ENGINEERING

The proposed design is depicted in Figure 1. We set the GaP waveguide thickness to be $H_{wg}$=600 nm based on the measurement of our purchased GaP epi wafers. A partially etched gap is carved out which splits the waveguide into two. The inner waveguide width is denoted as $W_{in}$ and the outer is $W_{out}$. The radius $R$ is referred to the inner ring resonator and is chosen to be 140 µm because it can support an FSR of about 100 GHz at 1550 nm. The waveguide is upper cladded with $SiO_2$ for protection.

The partially etched gap is a key enabling factor for the two-color frequency comb generation. For Kerr OFC generation, anomalous dispersion is necessary to balance the phase shift induced by the Kerr nonlinear process. The anomalous dispersion condition is satisfied by two different mechanisms in our design. At shorter wavelengths (~1.55 µm), the structure can be viewed as two coupled waveguides because of the existence of the gap, and the anomalous dispersion is manifested by the avoided mode crossing [21] between two fundamental modes in the adjacent waveguides. In our previous work [22], we theoretically proved the feasibility of using concentric ring structure to manipulate the GaP-OI resonator dispersion. At longer wavelengths (~3.1 µm), the anomalous dispersion is realized by using a higher order TE mode. Since the size of the gap is relatively small compared to the wavelength. The impact of the partially etched gap can be neglected, especially for the $TE_1$ mode that naturally has a field null in the center. Therefore, the size of the gap needs to be carefully chosen to enable anomalous dispersion at shorter wavelengths without disturbing the mode distribution for higher order modes at longer wavelength.

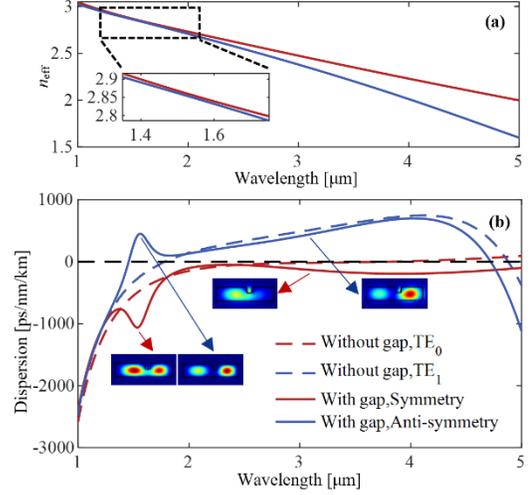

Figure 2 (a) The effective indices of the first two modes versus wavelength. The inset magnifies the spectral region of the avoided mode crossing. (b) The dispersion of the first two modes with/without a partially etched gap (solid/dashed lines). The blue line represents the anti-symmetric mode at shorter wavelength and $TE_1$ mode at longer wavelengths. The red line represents the symmetric mode at shorter wavelength and $TE_0$ mode at longer wavelength. Mode profiles are included as insets to illustrate the field distributions.

Our proposed structure has $W_{gap}$ = 410 nm, $H_{gap}$ = 250 nm, and the gap lateral position is optimized such that $W_{out}$ = 1080 nm and $W_{in}$ = 1500 nm. The effective indices ($n_{eff}$) and the dispersion profiles ($D$) of the first two modes are depicted in Figure 2. Around 1550 nm, the effective indices of the first two modes approach together due to mode hybridization. The mode hybridization is the phenomenon that two guided modes with similar mode indices in the waveguide are coupled with each other upon phase matching. The phase-matching condition is met when the round-trip optical path length (OPL, OPL=$2\pi R n_{eff}$) of the inner and outer rings are equal. Since the radius of the inner ring is smaller than the radius of the outer ring, the $n_{eff}$ of the inner ring should be slightly larger than that of the outer ring to enable round-trip phase matching. Close to the OPL matching wavelength, the two fundamental TE modes evolve into a pair of anti-symmetric mode and symmetric mode. The deviations in the $n_{eff}$, caused by mode coupling, alter the dispersion profiles of the two modes, making the anti-symmetric mode dispersion anomalous, as illustrated in Figure 2(b). The anti-symmetric mode gradually evolves into a $TE_1$ mode at longer wavelengths, whose geometric dispersion compensates for the material dispersion. Thus, anomalous dispersion is still maintained in the spectral range from 1.8 µm to 4.7 µm. The dispersion profiles of first two TE modes on the strip waveguide with the same dimension but without the partially etched gap are also added for comparison, as indicated by the dashed lines in Figure 2(b). Without the gap, there is no mode hybridization, and neither the $TE_0$ mode nor the $TE_1$ mode has anomalous dispersion around 1550 nm wavelengths.

To verify the fabrication variation tolerance of our proposed structure, we scrutinize the dispersion dependence on the partially etched gap dimension as well as the waveguide dimension, as illustrated in Figure 3.

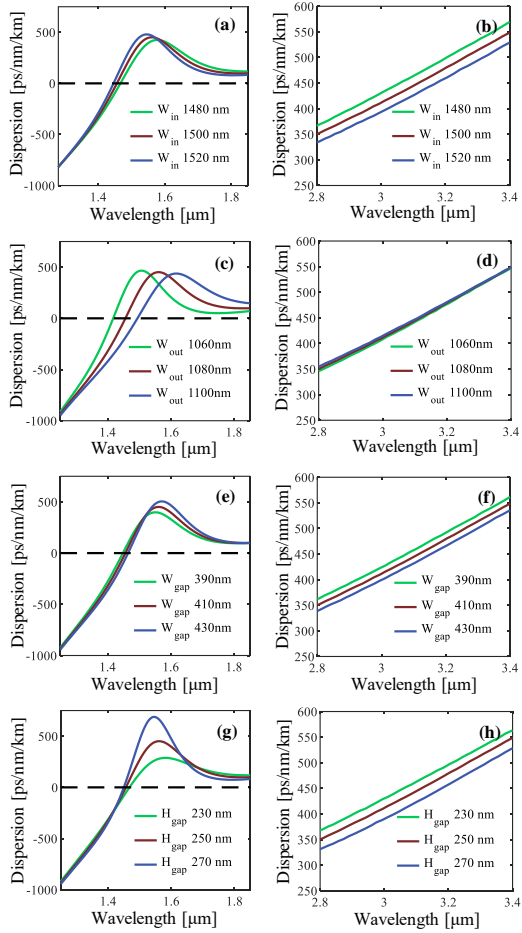

Figure 3 The dispersion profiles for anti-symmetric modes of different parameter sets of $W_{in}$, $W_{out}$, $W_{gap}$, and $H_{gap}$. (a), (c), (e), (g) correspond to the dispersion profile around 1550 nm wavelength. (b), (d), (f), (h) correspond to the dispersion profile around 3100 nm wavelength. Reference parameter set: $R = 140$ μm, $H_{wg}= 600$ nm, $W_{in} = 1500$ nm, $W_{out} = 1080$ nm, $W_{gap}= 410$ nm, and $H_{gap}= 250$ nm.

It can be seen that the anomalous dispersion profile is less sensitive to the $W_{in}$ and the $W_{gap}$ parameters around the 1550 nm wavelengths, as indicated in Figure 3(a) and Figure 3(e). These two parameters can be used to fine-tune the anomalous dispersion distribution. In contrast, $W_{out}$ and $H_{gap}$ are more effective in tuning the waveguide dispersion. There is a prominent red shift of the dispersion peak location when increasing $W_{out}$ (Figure 3(c)), whereas the $H_{gap}$ parameter has more influence on the dispersion peak amplitude shown in Figure 3(g). The dispersion sensitivity on $W_{out}$ and $H_{gap}$ can be utilized to coarsely adjust the anomalous dispersion distribution. These tuning behaviors can be explained by the coupled mode theory [23] and phase matching condition [21].

For the dispersion profiles around 3100 nm wavelengths, the amplitude of the anomalous dispersion decreases as the width of the inner waveguide increases (Figure 3(b)). By contrast, the width of the outer waveguide has a neglectable effect on the waveguide dispersion (Figure 3(d)). Either increasing the gap width or increasing the gap depth can lead to a decrease in anomalous dispersion amplitude, shown in Figure 3(f) and Figure 3(g). In general, the structure has more fabrication variation tolerance at 3.1 μm wavelength than at 1.55 μm wavelength, because the relative variations compared to the wavelength is smaller at longer wavelength.

### III. KERR FREQUENCY COMB GENERATION

We numerically investigate multicolor optical frequency comb generation in our optimized structure with the parameter set of $R = 140$ μm, $H_{wg}=600$ nm, $W_{in}=1500$ nm, $W_{out}=1080$ nm, $W_{gap}=410$ nm, and $H_{gap}=250$ nm. We assume a propagation loss of 1.76 dB/cm in the simulation which is reasonably within the reported values from literature [18], [24], [25]. With proper selections of pump power, laser detuning rate, two dissipative Kerr soliton frequency combs can be obtained individually around 1500 nm and 3100 nm spectral windows by solving the Lugiato-Lefever Equation [26], shown in Figure 4(b). The frequency combs have a tooth spacing of 101 GHz at 1550 nm wavelength and of 97 GHz at 3100 nm wavelength. The spectral profiles exhibit a $sech^2$ envelope in each spectral window, which indicates that temporally there is a single soliton circulates through the cavity.

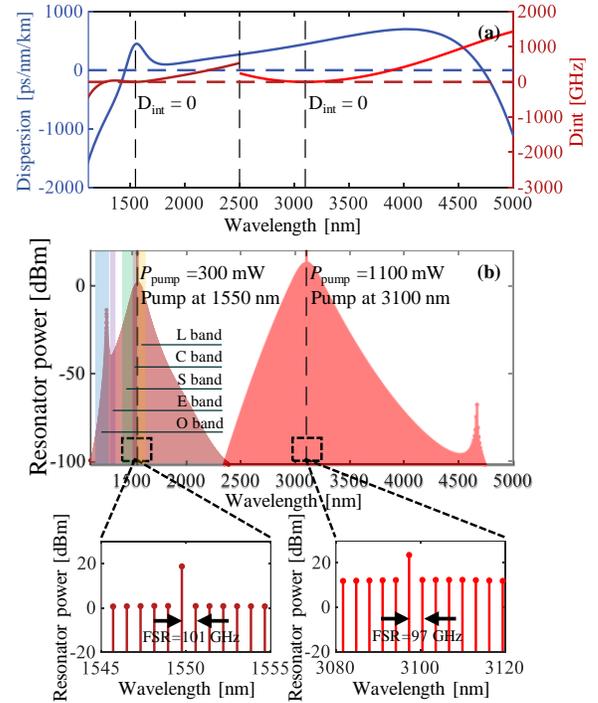

Figure 4 (a) Dispersion profiles of the anti-symmetric mode (blue, left y-axis) and integrated dispersion (red, right y-axis) of the anti-symmetric mode. The light blue areas highlight the anomalous dispersion regions of the antisymmetric mode. (b) The simulated frequency comb spectra inside the resonator. Each comb spectrum follows a $sech^2$ envelope centered at its pump wavelength. (c) The zoomed-in comb teeth around centered 1550 nm with a spacing of 101 GHz. The zoomed-in comb teeth centered 3100 nm with a spacing of 97 GHz.

## IV. CONCLUSION

In conclusion, we presented a dispersion engineering approach using a partially etched gap in the microring resonator to support multicolor frequency combs. The required anomalous dispersion was fulfilled by the anti-symmetric mode at 1550 nm and the $TE_1$ mode at 3100 nm wavelengths. We demonstrate this method on a 600 nm-thick GaP-OI resonator and found a 3.25 μm wide anomalous dispersion span, which could not be achieved on a solely strip waveguide without the gap. The fabrication variation tolerance of the optimized structure was digested, and structure exhibits better tolerance at longer wavelengths. Kerr soliton combs are theoretically predicted at the 1550 nm and the 3100 nm spectral windows. This proposed design offers more dispersion tuning knobs with the help of the partially etched gap, opening the possibility to achieve coherent multicolor frequency combs in the emerging GaP-OI integrated photonic platform.

## ACKNOWLEDGMENT

This work is supported by Guangdong Basic and Applied Basic Research Foundation under the Grant Number 2021B1515120074.